\documentclass[aip,apl,amsmath,amssymb,preprint,superscriptaddress]{revtex4}

\usepackage[version=3]{mhchem} 
\usepackage{graphicx}
\usepackage{dcolumn}
\usepackage{bm}
\usepackage{xr}
\usepackage{color}
\usepackage{amssymb,amsmath}

\newcommand{\unit}[1]{~\mathrm{#1}}

\begin{document}
\author{A.~C.~Betz}
\email{ab2106@cam.ac.uk}
\affiliation{Hitachi Cambridge Laboratory, J. J. Thomson Avenue, Cambridge CB3 0HE, United Kingdom}
\author{R.~Wacquez}
\affiliation{CEA/LETI-MINATEC, CEA-Grenoble, 17 rue des martyrs, F-38054 Grenoble, France}
\author{M.~Vinet}
\affiliation{CEA/LETI-MINATEC, CEA-Grenoble, 17 rue des martyrs, F-38054 Grenoble, France}
\author{X.~Jehl}
\affiliation{SPSMS, UMR-E CEA / UJF-Grenoble 1, INAC, 17 rue des Martyrs, 38054 Grenoble, France}
\author{A.~L.~Saraiva}
\affiliation{Instituto de Fisica, Universidade Federal do Rio de Janeiro, Caixa Postal 68528, 21941-972 Rio de Janeiro, Brazil}
\author{M.~Sanquer}
\affiliation{SPSMS, UMR-E CEA / UJF-Grenoble 1, INAC, 17 rue des Martyrs, 38054 Grenoble, France}
\author{A.~J.~Ferguson}
\affiliation{Cavendish Laboratory, University of Cambridge, CB3 0HE, United Kingdom}
\author{M.~F.~Gonzalez~-~Zalba}
\affiliation{Hitachi Cambridge Laboratory, J. J. Thomson Avenue, Cambridge CB3 0HE, United Kingdom}

\date{\today}

\title
{Dispersively detected Pauli Spin-Blockade in a Silicon Nanowire Field-Effect Transistor}


\begin{abstract}
We report the dispersive readout of the spin state of a double quantum dot formed at the corner states of a silicon nanowire field-effect transistor. Two face-to-face top-gate electrodes allow us to independently tune the charge occupation of the quantum dot system down to the few-electron limit. We measure the charge stability of the double quantum dot in DC transport as well as dispersively via in-situ gate-based radio frequency reflectometry, where one top-gate electrode is connected to a resonator. The latter removes the need for external charge sensors in quantum computing architectures and provides a compact way to readout the dispersive shift caused by changes in the quantum capacitance during interdot charge transitions. Here, we observe Pauli spin-blockade in the high-frequency response of the circuit at finite magnetic fields between singlet and triplet states. The blockade is lifted at higher magnetic fields when intra-dot triplet states become the ground state configuration. A lineshape analysis of the dispersive phase shift reveals furthermore an intradot valley-orbit splitting $\Delta_{vo}$ of $145\unit{\mu eV}$. Our results open up the possibility to operate compact CMOS technology as a singlet-triplet qubit and make split-gate silicon nanowire architectures an ideal candidate for the study of spin dynamics.
\end{abstract}
\maketitle

Since its first observation in electrical transport through a GaAs double quantum dot spin-blockade, or Pauli blockade, has been important for understanding and controlling the behaviour of artificial atoms \cite{Ono2002Science}. Spin-blockade has enabled the effect of the nuclear spin environment to be probed, showing that nuclear spins are responsible for mixing between the electron spin states \cite{Koppens2005Science,Johnson2005Nature}. Perhaps its biggest contribution is allowing read-out of the spin state of a double quantum dot in the singlet-triplet basis \cite{Petta2005,JohnsonPRB2005}. In such experiments a charge sensor will typically monitor the response of the double quantum dot charge to an external voltage pulse, yielding an averaged or single shot measurement of the spin state \cite{Petta2005,Barthel2009}. Spin-blockade has proved to be a generic tool for quantum dots made in materials other than GaAs, and most recently is of interest in silicon \cite{Weber2014,Nguyen2014,Yamahata2012,Lai2011,Liu2008} or silicon germanium \cite{Maune2012, Wu2014, Prance2012}, where a reduced nuclear spin environment leads to longer spin coherence times \cite{Veldhorst2014}. In addition it is possible to measure spin-blockade without a charge sensor and without direct electrical transport, by performing high-frequency capacitance measurements on the double quantum dot \cite{Petersson2010NL}. Removing the need for an external charge sensor makes such quantum capacitance or dispersive measurements attractive as it reduces the complexity of the gate architecture of the quantum dots. It also enables a double quantum dot to be interfaced with superconducting resonators \cite{Petersson2012}, a promising element for long-distance transfer of quantum information. 

In this Letter, we demonstrate the in-situ dispersive measurement of spin-blockade in the corner states of a double-gated silicon nanowire FET \cite{Voisin2014NL,Dupont-Ferrier2013} (SiNWFET) and describe the magnetic field dependence of the high-frequency response of the system. A double quantum dot (DQD) arises as a result of electrostatics in the square channel transistor geometry \cite{Gonzalez2015NC,Voisin2014NL}. Each dot can be tuned independently by a separate top-gate electrode, one of which is connected to a resonant circuit and provides the dispersive MHz reflectometry readout. We confirm the presence of one quantum dot per corner state and observe the double quantum dot's few-electron limit and interdot charge transitions. The latter manifest as additional capacitance contribution, readily captured by the dispersive gate sensor. We use this to selectively readout the spin state of a singlet-triplet charge transition under finite magnetic field. Furthermore, we observe lifting of the blockade at higher magnetic fields due to anti-crossing triplet states. Our work shows a compact way to measure spin-blockade in a device processed using standard industrial fabrication, paving the way towards fully CMOS-compatible quantum computing architectures \cite{Angus2007}.

\begin{figure*}[ht]
\centering
\includegraphics[width=\textwidth]{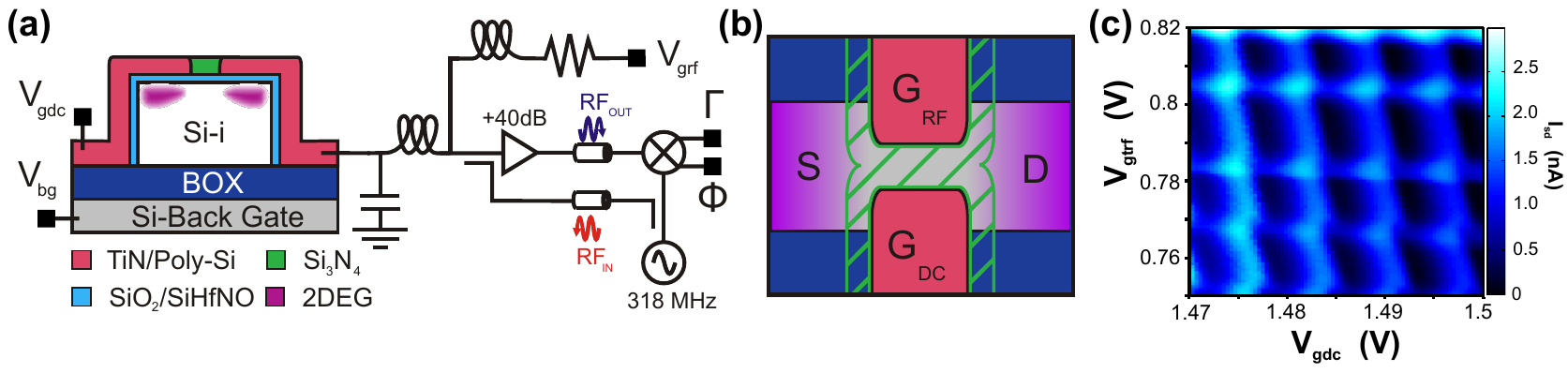}
\caption{Device geometry and measurement setup (a) Sketch of device cross-section perpendicular to transport direction and reflectometry setup. When the  top-gates are biased just below threshold a double quantum dot is formed at the top most corners. Gate $\mathrm{G_{RF}}$ is connected to resonant circuitry including a bias-tee and used to probe the high-frequency admittance of the DQD. (b) Top view sketch presenting the face-to-face top-gate design. Si$_3$N$_4$ spacers (hatched green) provide the doping gradient from the doped source/drain to the intrinsic channel and separate the top-gates $\mathrm{G_{DC}}$ and $\mathrm{G_{RF}}$. (c) DC charge stability diagram of the double quantum dot in the multi-electron regime at $V_{sd}=3\unit{mV}$ and $V_{bg}=-0.5\unit{V}$. The honeycomb pattern, typical for interacting quantum dots, indicates low cross-coupling capacitance.}\label{Fig1}
\end{figure*}

The device presented here is a fully depleted silicon-on-insulator (SOI) nanowire transistor \cite{Hofheinz2006APL,Roche2012APL}. Fig.\ref{Fig1}(a) shows a cross-section of the transistor perpendicular to the transport direction. It consists of an undoped Si (001) channel of thickness $t_{Si}=12\unit{nm}$ and width $W=100\unit{nm}$ on top of a $150\unit{nm}$ $\mathrm{SiO_2}$ buried oxide (BOX). The underlying Si wafer serves as global back-gate. The channel is furthermore locally controlled by two face-to-face top-gate electrodes (length $L=60\unit{nm}$), separated from the channel by $5\unit{nm}$ of $\mathrm{SiO_2}$ (Fig.\ref{Fig1}(a) and (b)) and at a distance of $70\unit{nm}$ from each other. The electrical isolation between them is provided by $\mathrm{Si_3 N_4}$ spacers, which extend $40\unit{nm}$ towards source and drain to prevent dopant diffusion from the highly doped contacts into the channel during fabrication (see Fig.\ref{Fig1}(b)).\newline
Measurements are taken via DC transport, recording the source-drain current $I_{sd}$, and via gate-based reflectometry readout \cite{Gonzalez2015NC,Colless2013PRL}, both carried out at the base temperature of a dilution refrigerator. A schematic of the reflectometry setup is shown in Fig.\ref{Fig1}(a): Gate electrode $\mathrm{G_{RF}}$ is coupled to a resonant LC circuit that consists of a surface mount inductor ($L=390\unit{nH}$) and the device's parasitic capacitance to ground ($C_p\simeq500\unit{fF}$). The DC voltage $V_{grf}$ across $G_{RF}$ is provided via an on-board bias-tee. We apply an RF tone of power $-88\unit{dBm}$ at the tank circuit's resonant frequency ($f_r=1/(2\pi\sqrt{LC_p})\simeq318\unit{MHz}$). Here, magnitude $\Gamma$ and phase $\Phi$ of the reflected signal are sensitive to changes in the device admittance and ultimately arise from excess power dissipation and susceptance changes, respectively \cite{Gonzalez2015NC,Colless2013PRL,Ciccarelli2011NJP,Chorley2012PRL}. In particular, the phase response $\Delta \Phi$ relates to an effective change in capacitance of the system, $\Delta C$, given by $\Delta\Phi\approx-\pi Q\Delta C/C_p$ \cite{Chorley2012PRL}. Here, Q is the quality factor of the resonator. We obtain $\Gamma$ and $\Phi$ from IQ-demodulation, after the signal has been amplified at low and room temperature.\newline
As displayed in Fig.\ref{Fig1}(a), the channel has a square cross-section and the top-gates cover two channel sides (or parts thereof) each. The electric field exerted by the gate electrodes is strongest at the top-most corner of the channel, where two gate faces meet. Charge accumulation in SiNWFETs occurs therefore first under the top-most corners, due to this so called corner effect \cite{Sellier2007APL}. The corner states are furthermore confined in SD-direction by the large spacers, resulting in a quantum dot under each corner \cite{Gonzalez2015NC,Voisin2014NL}, tunnel coupled to source and drain, as well as to the other dot.
We confirm the formation of the aforementioned double quantum dot in the DC transport measurement of Fig.\ref{Fig1}(c). It shows the source-drain current $I_{sd}$ as a function of both DC top-gate voltages $V_{gdc}$ and $V_{grf}$ in the multi-electron regime at $V_{sd}=3\unit{mV}$ and $V_{bg}=-0.5\unit{V}$. We observe the honeycomb diagram characteristic for coupled double quantum dots. The DQD is in parallel configuration, which allows us to observe transport not only at the triple points, as it is commonly seen in serial DQDs \cite{VanderWiel2002}, but also the individual transport lines of each dot. Conduction is however increased at the triple points, due to the increase in charge transport pathways. From the honeycomb diagram, we extract the gate voltage spacing $\Delta V_{gdc(grf)}=7\,(17)\unit{mV}$. Moreover, $V_{sd}-V_{gxx}$ maps of the individual dots, obtained with the respective other dot biased largely below threshold, indicate charging energies $E_C=3-5\unit{meV}$ and lever arms $\alpha=C_g/C_\Sigma\simeq 0.29$ ($\mathrm{G_{RF}}$) and $0.45$ ($\mathrm{G_{DC}}$) where $C_{\Sigma}$ is the total and $C_g$ the gate capacitance of the respective dot (See supplementary information). We note that the individual transport lines in Fig.\ref{Fig1}(c) only vary little with respect to the opposite gate voltage, which indicates low cross-coupling capacitances $C_c=1.4-2.0\unit{aF}$. Finally, from the shift produced in the vertical and horizontal lines due to the charging of an electron in the opposite dot we infer a mutual capacitance $C_m=5\unit{aF}$. The main feature here, from a quantum information processing point of view, is the possibility to individually and independently control two coupled quantum dots fabricated in an industry-standard CMOS transistor. 
\begin{figure*}[htp]
\centering
\includegraphics[width=\textwidth]{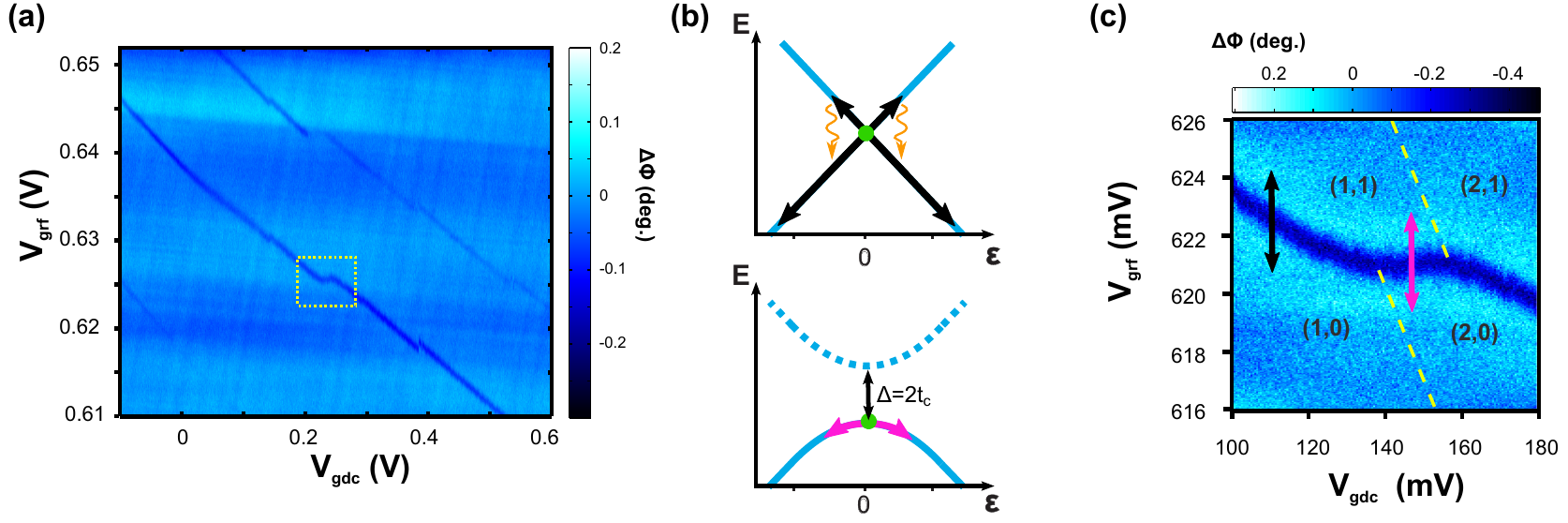}
\caption{Dispersive gate based RF readout in the few electron regime. (a) Stability diagram at $V_{sd}=0\unit{V}$ and $V_{bg}=-1\unit{V}$. Only lines originating from $QD_{RF}$ are detected by the gate sensor. The dashed square indicates the transition shown in (c). (b) RF readout model. Top: For lead-dot transitions the system is modelled by two crossing levels. The signal arises from a charge being cyclically driven across the charge instability at $V_0$ by the RF excitation $\Delta V_g$ (black arrows) and the subsequent relaxation (orange). Bottom: Inter-dot transitions are measurable due to the additional quantum capacitance that arises from the band curvature at the anti-crossing. $\epsilon$ is the inter-dot detuning. (c) Close-up of inter-dot transition. The arrows indicate the two measurement modes shown in (b). The dashed lines are an approximation of the lines of $QD_{DC}$.}\label{Fig2}
\end{figure*}


We turn now to the high-frequency response of the system. Fig.\ref{Fig2}(a) shows the dispersive measurement of the DQD's charge stability in the few-electron regime, obtained via gate-based RF detection where we plot the change in phase $\Delta\Phi$. Only electron transitions involving the dot $QD_{RF}$ under gate $G_{RF}$ are visible now (dark blue diagonal traces), since the resonant circuit is coupled strongly to $QD_{RF}$ via $G_{RF}$ and only weakly to the other dot $QD_{DC}$. The signal corresponds to cyclic single-charge tunnelling between the source or drain and $QD_{RF}$ driven by the MHz tone applied to the resonator. In this case, the system can be modelled as inelastic charge transitions in a fast-driven two-level system \cite{Ciccarelli2011NJP,Gonzalez2015NC}, as shown in the top panel of Fig.\ref{Fig2}(b). Electrons tunnelling out-of-phase with the RF-drive lead to an additional tunnelling capacitance contribution, which manifests as a phase change in the resonator's response. Besides simplifying the architecture, the advantage of this charge readout technique is that no direct transport is necessary, only the possibility to cyclically exchange electrons between a reservoir and the probed quantum dot.
As in the multi-electron regime of Fig.\ref{Fig1}(c), we find very little cross-coupling between the gates (note the difference in x and y scales) and a similar voltage spacing $\Delta V_{grf}$. The kinks observable in $QD_{RF}$'s transition lines indicate coupling to charge transitions separate from $QD_{RF}$. We attribute this to $QD_{DC}$, which in this low voltage regime seems to be disordered due to surface roughness \cite{Voisin2014NL}: The voltage spacing $\Delta V_{gdc}$ has increased from the multi-electron regime and has become more irregular.
%
Fig.\ref{Fig2}(c) shows a zoom into the region of one of the first transitions into $QD_{DC}$. The line corresponding to the loading of an electron into $QD_{RF}$ (black arrow) is interjected by a ridge corresponding to interdot charge exchange. Yellow dashed lines furthermore approximate the position of lead-$QD_{DC}$ transitions. We attribute the presence of signal along the interdot ridge to a mechanism different to tunnelling capacitance described previously: Tunnel coupling $t$ between $QD_{RF}$ and $QD_{DC}$ causes the DQD energy bands to hybridise with energies $E_{\pm}=\pm\sqrt{\epsilon^2+(2t)^2}/2$, as displayed in the lower panel of Fig.\ref{Fig2}(b) as a function of detuning $\epsilon$. For $f_r<t/h$, there is no tunnelling capacitance since the charge remains in the ground state while the RF drive cycles it across the interdot line. However, an additional capacitance contribution now arises from the band curvature, the so called quantum capacitance \cite{Sillanpaa2005,Duty2005}
\begin{equation}\label{EqCq}
C_q^\pm=-(e\alpha)^2\frac{\partial^2 E_\pm}{\partial\epsilon^2}
\end{equation} 
where $e$ is the charge of the electron and $\alpha$ the resonator-DQD coupling given by the gate lever arm \cite{Petersson2010NL}. $C_q$'s capacitance contribution is maximum at $\epsilon=0$ and positive (negative) for the ground (excited) state. The capacitance change due to $C_q$ is picked up dispersively by the gate sensor (magenta arrow in Fig.\ref{Fig2}(c)). This renders the interdot transition visible and makes it possible to not only probe the interaction of $QD_{RF}$ with a charge reservoir via tunneling capacitance, but also the hybridised double quantum dot with the compact dispersive gate-sensor.
Note that we estimated electron occupation numbers in the form ($QD_{DC}$, $QD_{RF}$) in Fig.\ref{Fig2}(c). The absence of further transition kinks towards lower $V_{gdc}$ in Fig.\ref{Fig2}(a) indicates that we likely observe the first electrons into $QD_{DC}$. For $QD_{RF}$, although we have labelled the interdot transition (1,1) - (2,0) as a guide, the electron number corresponds to the valence occupancy. However, it becomes clear below that this transition is also amongst the last few electrons of $QD_{RF}$.

\begin{figure}[htp]
\centering
\includegraphics[width=0.7\textwidth]{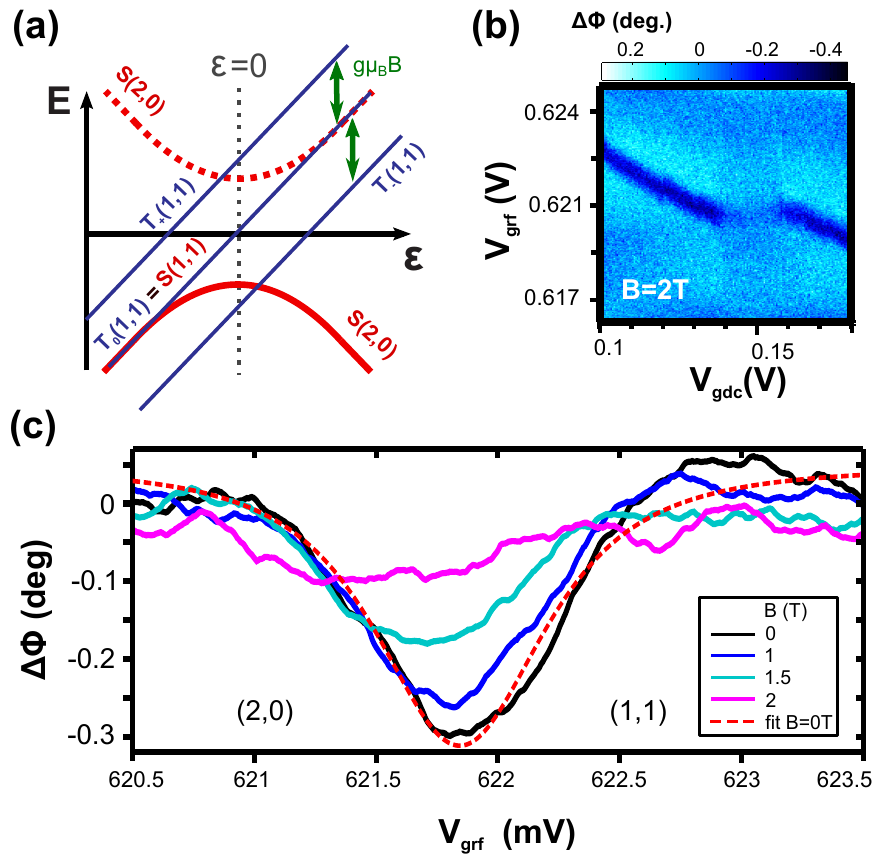}
\caption{Dispersive spin-blockade readout. (a) Schematic of the energy bands in the DQD, including singlet and triplet lines. The triplet configuration separates into $T_+$, $T_0$ and $T_-$ spaced at $\Delta E=g\mu_B B$. (b) Close-up of interdot transition of Fig.\ref{Fig2}(c) for finite magnetic field. Relaxation into $T_-$ at $\epsilon<0$ reduces the interdot phase signal. (c) Traces $\Delta\Phi\,(V_{grf})$ for different magnetic fields taken at $V_{gdc}=0.15\unit{V}$. The quantum capacitance signal decreases with increasing magnetic field due to the aforementioned relaxation.}\label{Fig3}
\end{figure}

We now turn to the investigation of spin-related effects in the DQD and in particular in the interdot transition of Fig.\ref{Fig2}(c). In the following, we demonstrate that Fig.\ref{Fig2}(c) depicts an even parity, spin-dependent transition. The reason for this becomes clear when we apply a magnetic field: Fig.\ref{Fig3}(b) shows the interdot transition in question at $B=2\unit{T}$. We find that while the tunnelling capacitance signal of the lead-dot transitions remains unchanged, the interdot ridge vanishes with increasing magnetic field, as can be seen in Fig.\ref{Fig3}(c). Here, we show traces across the interdot ridge at $V_{gdc}=0.15\unit{V}$ for different magnetic fields. We attribute loss of interdot signal to Pauli spin-blockade between a joint singlet $S(2,0)$ and a separated triplet $T(1,1)$, as shown in Fig.\ref{Fig3}(a). At zero magnetic field, the ground state of the system at zero detuning is singlet and the resonator is sensitive to its finite quantum capacitance as depicted in Fig.\ref{Fig2}(c). For $B\neq 0$ however, the triplet state is Zeeman-split into $T_0$ and $T_\pm$, with the latter spaced at $\Delta E=\pm g\mu_B B$ from $T_0$ \cite{Stephanenko2012PRB}. Here, $g$ is the electron spin g-factor and $\mu_B$ the Bohr magneton. When $g\mu_B B>t$, the $T_-(1,1)$ state becomes the ground state at zero detuning. As a result interdot phase signal decreases, since $T_-(1,1)$ is linear in $\epsilon$, i.e. has no capacitive contribution due to its lack of band curvature (see Eq.\ref{EqCq}), as observed in Fig.\ref{Fig3}(c). We furthermore analyse the $B=0\unit{T}$ trace quantitatively by fitting it with the equation
\begin{equation}\label{EqCq2}
\Delta\Phi \propto e^2\frac{\left(2t\right)^2}{\left[ \alpha^2 (V_{grf}-V_{grf}^0)^2+\left(2t\right)^2\right]^{(3/2)}}
\end{equation}
obtained from Eq.(\ref{EqCq}), the hybridised DQD energy branches and $\Delta\Phi\propto\Delta C$. The result is indicated by the dashed line in Fig.\ref{Fig3}(c).
Owing to the low cross-capacitance we use the lever arm $\alpha\simeq 0.29$ found previously for the conversion between gate voltage and detuning energy \cite{VanderWiel2002}, and find a tunnel coupling $t\simeq 80\unit{\mu eV}$ for the transition studied here. 
In total, the above results demonstrate that it is possible to read-out the spin state of individual electrons in our SiNWFET geometry, one of the essential ingredients to implement an interacting spin qubit in standard CMOS technology. 

\begin{figure}[htp]
\centering
\includegraphics[width=0.7\textwidth]{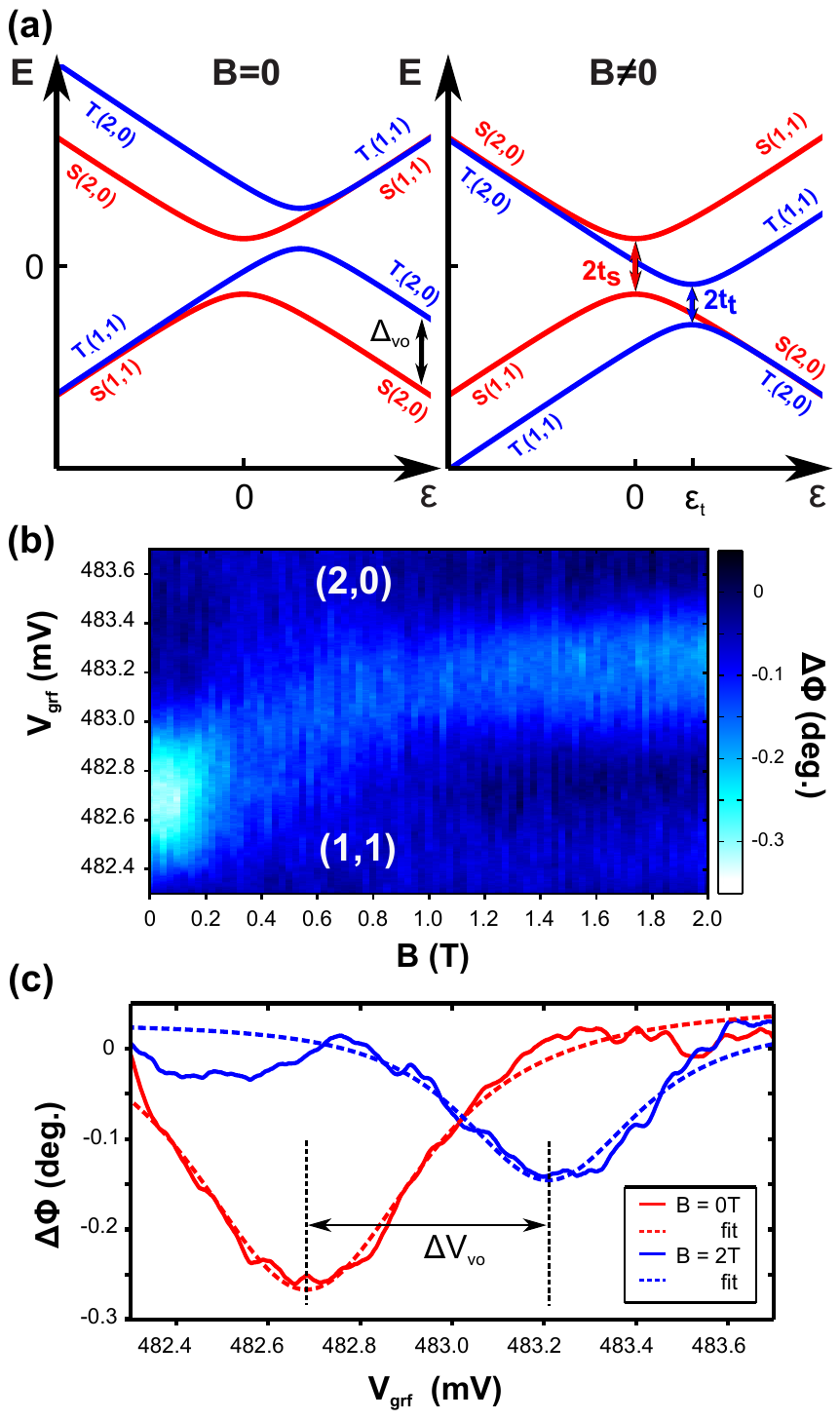}
\caption{spin-blockade lifting and triplet detection. (a) Schematic of lowest singlet and triplet branches at zero and finite magnetic field. (b)  Phase signal as function of $V_{grf}$ and magnetic field. (c) Phase as function of $V_{grf}$ for singlet (red) and triplet (blue). Dashed lines are fits to the data following Eq.(\ref{EqCq2}).}\label{Fig4}
\end{figure}

Finally, we explore spin-blockade lifting via triplet tunnelling. The band diagram discussed above only takes into account separated triplets, as $T(2,0)$ states are generally elevated in energy due to intradot valley-orbit coupling in silicon \cite{Lim2011,Saraiva2010,Culcer2009}. The $T(1,1)$ and $T(2,0)$ states hybridise similarly to the singlet branch and a second anticrossing develops as sketched on the left of Fig.\ref{Fig4}(a) for zero magnetic field. This anti-crossing appears at $\epsilon\neq0$ when the singlet and triplet branches are separated by the valley-orbit splitting $\Delta_{vo}$. The tunnel couplings are given by $t_s$ and $t_t$ for singlet and triplet anti-crossing, respectively. Analogous to the $T(1,1)$ states, the $(2,0)$ triplets Zeeman-split into $T_0(2,0)$, $T_+(2,0)$ and $T_-(2,0)$ and the $T_-$ branch becomes the DQD's ground state when $g\mu_BB>\Delta_{vo}$. For simplicity we only show the singlet and $T_-$ branch in Fig.\ref{Fig4}(a).\newline
We have investigated this scenario in the measurement shown in Fig.\ref{Fig4}(b), where we perform a magnetic field sweep across an even-parity transition lying further below in voltage $V_{grf}$ than the one studied in Fig.\ref{Fig2}(c) and Fig.\ref{Fig3}. At low magnetic fields we observe in Fig.\ref{Fig4}(b) the decrease in dispersive signal at $V_{grf}\simeq 482.7\unit{mV}$, i.e. $\epsilon=0$, due to relaxation from singlet to triplet. With increasing field however, a second dispersive signal arises, shifted upwards in $V_{grf}$, i.e. positive detuning. This represents the emerging of the triplet anti-crossing: Due to their newly acquired band curvature the $T_-$ states are now detectable by the dispersive gate sensor and can be distinguished from the singlet branch due to the detuning shift.\newline
Further to the qualitative understanding presented so far, we now analyse the singlet-triplet data quantitatively. To this end, we plot in Fig.\ref{Fig4}(c) traces $\Delta\Phi (V_{grf})$ for $B=0\unit{T}$ and $2\unit{T}$, which correspond to the two scenarios of Fig.\ref{Fig4}(a). At $B=0\unit{T}$, the ground state is predominantly singlet, whereas at $B=2\unit{T}$ the triplet dominates. Fitting with Eq.(\ref{EqCq2}), we find a singlet (triplet) tunnel coupling $t_{s(t)}\simeq 55\,(40)\unit{\mu eV}$ assuming, as previously, a lever arm $\alpha=0.29$. In order to fit the triplet data, we shift the detuning to $\epsilon_t=\epsilon-\Delta_{vo}$, which furthermore allows us to infer the valley-orbit (VO) splitting $\Delta_{vo}$ from the voltage difference $\Delta V_{vo}$ between singlet and triplet lineshapes. Converting to energy via the lever arm, we obtain thus $\Delta_{vo}=145\unit{\mu eV}$.
We confirm that the slanted nature of the corner dots, which face the (001), (010) crystal directions does not have a significant impact on the VO splitting since our results agree well with theoretical predictions for the valley splitting (0.1-0.3~meV) \cite{Saraiva2011B} and previously measured values in Si (0.1~meV) \cite{Lim2011,Jiang2013} and SiGe QDs (0.12-0.27~meV) \cite{Borselli2011}.

In conclusion, we have reported the dispersive readout of the spin state of a silicon nanowire corner state double quantum dot making use of the Pauli spin-blockade. By using a fully-based CMOS device consisting of two face-to-face top gates we have achieved independent control over the two quantum dots at the corner states of a Si nanowire transistor and by means of dispersive gate readout we have observed the double quantum dot's few electron regime. Spin-blockade manifests as a decrease in dispersive signal in a few-electron interdot transition of even parity. We have furthermore presented a scenario where singlet and triplet branches can be discerned by means of magnetic field studies. Our results demonstrate that compact CMOS-based architectures are suitable to implement spin qubits. 
Ultimately, this fully industrial approach opens a window to larger-scale qubit architectures.


The authors thank D.A. Williams and A. Esmail for fruitful discussion. The research leading to these results has been supported by the European Community's seventh Framework under the Grant Agreement No. 214989. The samples presented in this work were designed and fabricated by the AFSID project partners (www.afsid.eu). A.J. Ferguson acknowledges support from EPSRC (EP/K027018/1) and from his Hitachi research fellowship.



\end{document}